\documentstyle[12pt]{article}
\newcommand\be{\begin{equation}}
\newcommand\ee{\end{equation}}
\newcommand\bea{\begin{eqnarray}}
\newcommand\eea{\end{eqnarray}}

\newcommand\Tr{{\rm{Tr\,}}}
\newcommand\tr{{\rm{tr\,}}}

\newcommand\e{{\rm{e}}}

    \def\newpic#1{%
    \def\emline##1##2##3##4##5##6{%
       \put(##1,##2){\special{em:point #1##3}}%
       \put(##4,##5){\special{em:point #1##6}}%
       \special{em:line #1##3,#1##6}}}
 \newpic{}

\begin{document}

\begin{center}
{
\bf 
ONE-LOOP EFFECTIVE ACTION FOR THE EXTENDED\\ SPINOR
ELECTRODYNAMICS WITH VIOLATION OF\\ LORENTZ AND CPT SYMMETRY}

\bigskip
Yu.A.~Sitenko
\footnote
{e-mail:  yusitenko@bitp.kiev.ua }

\medskip
 {\it Center for Theoretical Physics, Massachusetts Institute of Technology,}\\ 
\medskip
 {\it   Cambridge, MA 02139-4307, USA}\\ 
\medskip
 {\it and}\\ 
\medskip
 {\it Bogolyubov Institute for Theoretical Physics, National Academy of Sciences,}\\
 \medskip
 {\it Kyiv, 03143, Ukraine}
\end{center}

\bigskip
\begin{abstract}
\begin{sloppypar} 
If violation of Lorentz and CPT symmetry is introduced into the fermion 
sector of conventional quantum electrodynamics, then the Chern-Simons term 
is radiatively induced with finite nonzero coefficient, as well as the Maxwell 
term is with logarithmically divergent one. The heat kernel expansion and the 
proper time methods are used to determine the effective action in the one-loop 
approximation unambiguously.
\end{sloppypar} 
\end{abstract}


\noindent
PACS numbers: 11.10.Gh, 11.30.Cp, 11.30.Er, 11.15.Tk

\noindent
Keywords: Chern-Simons invariant; Effective action; Heat kernel; 
Fock-Schwinger method
\thispagestyle{empty}


\vspace{1cm} 

Lorentz and CPT symmetry of conventional Maxwell electrodynamics is
  destroyed by inclusion a Chern-Simons term \cite{Jao} into the
  action \cite{Ca, Co1, Co2}
  \be
 \Gamma = -{1\over4}\int d^4x\,F^{\mu \nu}(x)F_{\mu\nu}(x)-
 k^\mu \, \int d^4x\,A^\nu(x)\widetilde{F}_{\mu\nu}(x),
 \ee
 where
 $F_{\mu\nu}=\partial_\mu A_\nu-\partial_\nu A_\mu$ is the
 electromagnetic field strength,
 $\widetilde{F}^{\mu\nu}={1\over2}
 \varepsilon^{\mu\nu\rho\omega}F_{\rho\omega}$
 is its dual, and metric of the Minkowski space-time is chosen as
 $g_{\mu\nu}={\rm diag}(1,-1,-1,-1)$. Gauge invariance of action (1)
 and of the corresponding equations of motion is maintained, provided
 that $ k^\mu$ is a constant 4-vector. This vector selects a preferred
 direction in space-time, thus violating both Lorentz invariance and
 discrete CPT symmetry. Observation of distant galaxies puts a
 stringent bound on the value of $k^\mu$: it should effectively vanish
 \cite{Go, Ja1}.

 On the other hand, a Lorentz- and CPT-violating extension of spinor
 electrodynamics can be proposed by adding an appropriate
 axial-vector interaction into the lagrangian
 \be
 L(x)=\bar{\psi}(x)\biggl(i\widehat{\partial}-e\widehat{A}(x)+
 \widehat{b}\gamma^5-m\biggr)
 \psi(x),
 \ee
 where $\widehat{\partial}=\gamma^\mu\partial_\mu$,
 $[\gamma^\mu,\gamma^\nu]_+=2g^{\mu\nu}$,
 $\gamma^5=-i\gamma^0\gamma^1\gamma^2\gamma^3$, and $b^\mu$ is a
 constant 4-vector. Then classical action (1) attains a quantum
 correction, resulting from integrating out the spinor degrees of
 freedom, which in the one-loop approximation reads:
 \be
 \Gamma^{\rm quant}=-i\ln\biggl\{\int d\bar{\psi}\,
 d\psi\, \exp[i\int d^4x\,L(x)]\biggr\}=-i\Tr\ln\biggl(i\widehat{\partial}-
 e\widehat{A}(x)+\widehat{b}\gamma^5-m\biggr),
 \ee
 where $\Tr U=\int d^4x\, \tr\langle x|U|x\rangle$ is the trace of an operator 
 in functional space, and the trace over spinor indices is denoted by $\tr$.

Natural questions arize: does the interaction of quantized spinor fields, as 
given by $L(x)$ (2), results in the emergence of the effective Chern-Simons 
($A\widetilde{F}$) term? and, if so, whether the observed smallness of the 
coefficient before the $A\widetilde{F}$ term should be translated into a 
restriction on the parameter of Lorentz and CPT symmetry violation in the 
fermionic sector ($b^\mu$)? The answers to this questions, as is known from 
the literature, are quite different, even controversial. Whereas the authors of 
Refs.\cite{CG, Bo} insist that the effective $A\widetilde{F}$ term is not induced 
by radiative corrections, the authors of Refs.\cite{Ja2, WC, JC1, PV, Na} 
get the radiatively induced $A\widetilde{F}$ term with coefficient 
${{-3e^2}\over{16\pi^2}} b^\mu$, and the authors of Refs.\cite{LC, MC, JC3} get 
this term with coefficient ${{-e^2}\over{8\pi^2}} b^\mu$. Meanwhile, it is also 
claimed that the $A\widetilde{F}$ term is induced with finite but ambiguous
coefficient \cite{Co2, Ja2, Ja3, JC2, Pe}. 

With the purpose of clarifying the situation, in the present Letter, a 
comprehensive analysis of one-loop effective action (3) will be carried out 
in the framework of covariant nonperturbative formalism involving the heat 
kernel expansion \cite{De,Se,Gi} and the Fock-Schwinger proper 
time \cite{Fo,Sch} methods. We shall find out that the $A\widetilde{F}$ term 
is radiatively induced with coefficient ${{-e^2}\over{4\pi^2}} b^\mu$. 
 
Taking a functional derivative of $\Gamma^{\rm quant}$ (3), let us define 
current
 \be
J^\mu(x)= - {\delta\Gamma^{\rm quant}\over\delta A_\mu(x)}= - ie \, \tr
\gamma^\mu \langle x|\biggl(i\widehat{\partial}
-e\widehat{A}(x)+\widehat{b}\gamma^5 - m\biggr)^{-1}
|x\rangle,
\ee
which can be presented in the form 
\bea
J^\mu(x) &=& ie \, \tr\gamma^\mu \langle
x|\biggl(i\widehat{\partial}-e\widehat{A}(x)-\widehat{b}\gamma^5+
m\biggr)\times \nonumber\\
&&\times \biggl[\biggl(-i\widehat{\partial}+e\widehat{A}(x)-\widehat{b}\gamma^5
+m\biggr) \biggl(i\widehat{\partial}-e\widehat{A}(x)-\widehat{b}\gamma^5+
m\biggr)\biggr]^{-1}|x\rangle =\nonumber\\
&&=ie\tr \gamma^\mu\langle
x|(\widehat{\pi}+m)(H^2+m^2)^{-1}|x\rangle ,
\eea
where
\be
H^2=-\pi^\rho(x)\pi_\rho(x)+{1\over2}\sigma^{\omega\rho}eF_{\omega\rho}(x)+
2m\gamma^5\widehat{b}
\ee
and
\be
\pi_\mu(x)=i\partial_\mu-eA_\mu(x)-b_\mu \gamma^5 \,  , \quad
[\pi_\mu(x),\pi_\nu(x)]_-=-ieF_{\mu\nu}(x)\, ,\quad
\sigma^{\mu\nu}={i\over2}[\gamma^\mu,\gamma^\nu]_-\, \,.
\ee 

Expressions (4) and (5) can be regarded as purely formal and, strictly
speaking, meaningless: they are ill-defined, suffering from infinite
divergencies. The well-defined quantities are nondiagonal $(x'\neq x)$
matrix elements of operators in these expressions. This suggests a
point-splitting procedure to regularize divergencies, i.e. computing
first the nondiagonal matrix elements and then tending $x'\to x$.
Another procedure is to modify the operator by multiplying it with a suitable
operator that smoothens singular behaviour of matrix elements as
$x'\to x$. In this Letter we shall obtain results  in the framework of 
both of these two regularization procedures.

Taking an exponential function of the denominator in the capacity of 
smoothening operator (which is reminiscent of Fujikawa's computation of 
axial anomaly \cite{Fu}), we define regularized current as
\be
J^\mu(x)=ie \, \tr\gamma^\mu\langle
x|(\widehat{\pi}+m)(H^2+m^2)^{-1}\exp[-t(H^2+m^2)]|x\rangle, 
\ee
where $t>0$ plays a role of a regularization parameter.
Upon the Wick rotation of the time axis $(x^0=-ix^4)$, operator
$H^2(6)$ becomes a positive definite elliptic operator. Our  aim is to
find out an asymptotic expansion of quantity (8) as $t\to
0_+ $; then divergent at $t\to 0_+$ terms will represent the
divergencies in current $J^\mu$.

The asymptotic expansion of the diagonal matrix element of the
exponential function of an elliptic second-order differential 
operator has the form
\be
\langle x|\e^{-t(H^2+m^2)}|x\rangle\mathop{\simeq}\limits_{t\to
0_+}\sum_{l=0}^{\infty}E_l(x|H^2+m^2)\, t^{-{d\over2}+l},
\ee
where summation runs over nonnegative integers and $d$ is the
dimension of the base space. This expansion (known as the asymptotic
heat kernel expansion) is extensively studied and intensely employed
in various areas of theoretical and mathematical physics. Coefficients
$E_l$ (known as De Witt-Seeley-Gilkey coefficients)
\cite{De,Se,Gi}) are endomorphisms of a fiber bundle, they are local
covariant quantities composed from the coefficient functions of
operator $H^2$, curvatures of the bundle and its base, and their
covariant derivatives. The asymptotic expansion of diagonal matrix
element
$$
\langle x|(H^2+m^2)^\alpha\e^{-t(H^2+m^2)}|x\rangle
$$
with an arbitrary real-valued $\alpha$ was obtained in Ref.\cite{Si1} 
where the expansion coefficients were expressed in terms of
$E_l(x|H^2+m^2)$ and $E_l(x|H^2)$. For our purpose  it suffices to use
the result (Eq.(26) in Ref.\cite{Si1}) for particular values
$\alpha=-1$ and $d=4$:
\bea
\langle
x|(H^2+m^2)^{-1}\e^{-t(H^2+m^2)}|x\rangle&\mathop{\simeq}\limits_{t\to0_+}&
i\biggl\{ E_0t^{-1}-E_1(x|H^2+m^2)
[\ln(tm^2)+\gamma]+\nonumber\\
&&+\sum_{l=2}^{\infty}E_l(x|H^2)m^{2-2l}(l-2)!\biggr\},
\eea
where factor $i$ emerges after a formal return back to Minkowski
space-time $(x^4=ix^0)$  and terms of order $t$, $t\ln t$ and higher
are omitted; $\gamma$ is the Euler constant.
Since $E_0=(4\pi)^{-2}$, the term with leading divergence $(t^{-1})$
does not contribute to $J^\mu$ (8). For a general operator $H^2$ in the form
\be
H^2=-\pi^\rho\pi_\rho+X,
\ee
the next coefficients in expansion (10) take form
\be
E_1(x|H^2+m^2)=-{1\over(4\pi)^2}\biggl(X+m^2\biggr).
\ee
\be
E_2(x|H^2)={1\over2(4\pi)^2}\biggl(X^2+{1\over3}\pi^\rho\pi_\rho X +
{1\over6}[\pi_\omega,\pi_\rho]_-[\pi^\omega,\pi^\rho]_-\biggr).
\ee
In the case of $H^2$ in the form of Eq.(6), one has
\be
X={1\over2}\sigma^{\omega\rho}eF_{\omega\rho}+2m\gamma^5\widehat{b}.
\ee

Decomposing $J^\mu$ (8) into two parts,  
\be
J^\mu(x)=J_1^\mu(x)+J_2^\mu(x),
\ee
where 
\be
J_1^\mu(x)=ie \, \tr\gamma^\mu\langle
x|\widehat{\pi}(H^2+m^2)^{-1}\exp[-t(H^2+m^2)]|x\rangle 
\ee
and
\be
J_2^\mu(x)=iem \, \tr\gamma^\mu\langle
x|(H^2+m^2)^{-1}\exp[-t(H^2+m^2)]|x\rangle, 
\ee
one can note that the $E_1$ coefficient contributes only to $J_1^\mu$ (16). 
Thus $J_2^\mu$ (17) is finite in the limit $t\to 0_+$. Computing the
contribution of the $E_2$ coefficient to Eq.(17) by taking an appropriate 
trace over $\gamma$-matrices, we find that only crossed terms in the square 
of $X$ (14) are relevant. Thus we get expression
\be
J^\mu_2(x)= - {e^2\over2\pi^2} \,\widetilde{F}^{\mu\nu}(x)b_\nu,
\ee
which yields the CPT-odd part of the effective action: 
\be
\Gamma^{\rm quant}_{\rm odd} = - \int d^4x\,A_\mu(x) J^\mu_2(x) 
= - {e^2\over4\pi^2}\,b^\mu\,\int d^4x\,A^\nu(x)\widetilde{F}_{\mu\nu}(x).
\ee

In order to compute current $J^\mu_1$ (16), we use the relation obtained 
in Ref.\cite{Si2} (see Eq.(31) there), which
after a formal return to the Minkowski space-time takes form 
\be
\langle x|\pi_\nu\e^{-t(H^2+m^2)}|x\rangle
\mathop{\simeq}\limits_{t\to0_+}
-i{t^{1-{d\over2}}\over2(4\pi)^{d/2}}\biggl(
\pi_\nu X+{1\over3}\pi^\rho
[\pi_\rho,\pi_\nu]_- \biggr)+O(t^{2-{d\over2}}).
\ee
Employing the results of Ref.\cite{Si1}, we get in the $d=4$ case
\be
\langle x|\pi_\nu(H^2+m^2)^{-1}\e^{-t(H^2+m^2)}|x\rangle
\mathop{\simeq}\limits_{t\to0_+}
i{\ln(tm^2)+\gamma\over2(4\pi)^2}\biggl(\pi_\nu X+{1\over3}\pi^\rho 
[\pi_\rho,\pi_\nu]_-\biggr),
\ee
where finite terms forming a series in inverse mass squared
($m^{2-2l}$,\ \ $l\geq2$) are omitted. It is now straightforward to
obtain expression
\be
J_1^\mu(x)={e^2\over12\pi^2} \, [\ln(tm^2)+\gamma] \, \partial_\nu
F^{\nu\mu}(x),
\ee
which yields the CPT-even part of the effective action:
\be
\Gamma^{\rm quant}_{\rm even} = - \int d^4x\,A_\mu(x) J^\mu_1(x)
={e^2\over48\pi^2} \, [\ln(tm^2)+\gamma] \, \int
d^4x\,F^{\mu\nu}(x)F_{\mu\nu}(x) \, .
\ee

Certainly, Eq.(23) is of no surprise, presenting the well-known
phenomenon of charge renormalization in the one-loop approximation
(for details see below). However, our intention was to compute both
$\Gamma^{\rm quant}_{\rm even}$
and
$\Gamma^{\rm quant}_{\rm odd}$
consistently, in the framework of the same regularization scheme. What
we have found is that $\Gamma^{\rm quant}_{\rm even}$ is logarithmically 
divergent and $\Gamma^{\rm quant}_{\rm odd}$ is finite; the coefficient 
before the Chern-Simons ($A\widetilde{F}$) term, as well as the coefficient 
before the logarithmically divergent part of the Maxwell ($F^2$) term, is 
determined unambiguously. 

The same result is maintained in the framework of the point-splitting 
regularization with the use of the Fock-Schwinger proper time method 
\cite{Fo,Sch}. The nondiagonal matrix element of $(H^2+m^2)^{-1}$ 
is presented in the form
\be
\langle
x|(H^2+m^2)^{-1}|x'\rangle = i \int\limits_{-\infty}^{0}
d\tau  \,\tr \langle x|\e^{i\tau (H^2+m^2)}|x'\rangle \, ,
\ee
where it is implied that the mass squared entails a small negative imaginary 
part, $m^2\to m^2 - i\epsilon$. Derivation of the expression for the 
integrand in Eq.(24) in the case of vanishing $b_\mu$ and constant uniform 
$F_{\mu\nu}$ can be found in a textbook (see, e.g., Ref.\cite{Itz}). 
Following exactly the same way, we have derived the expression in the 
case of nonvanishing constant $b_\mu$ and $F_{\mu\nu}$:
\bea
\hspace{-.5cm}\langle
x|(H^2+m^2)^{-1}|x'\rangle&\!\!\!\!=&\!\!\!\!{1\over(4\pi)^2}\exp\biggl(
-ie\int\limits_{x'}^{x}d\xi^\nu A_\nu - i\gamma^5 b_\nu x^\nu\biggr)
\int\limits_{-\infty}^{0}
{d\tau \over {\tau}^2}\biggl[\det\biggl({\sinh \tau eF \over 
\tau eF}\biggr)_{\rho\rho'}\biggr]^{-{1\over2}}
\times\nonumber\\
&&\hspace{-.5cm}\times
\exp\biggl\{ {i\over4}(x-x')^\omega (eF\coth \tau eF)_{\omega\omega'}
(x-x')^{\omega'} + i\tau \biggl({1\over2}\sigma^{\omega\omega'} 
eF_{\omega\omega'} + \nonumber\\
&&\hspace{-.5cm} + 2m\gamma^5\widehat{b} + m^2\biggr) + 2m\widehat{b} b^\omega 
[(eF)^{-1} + \tau (1 - \coth \tau eF)]_{\omega\omega'}(x-x')^{\omega'} 
-\nonumber\\
&&\hspace{-.5cm} - 4i\tau m^2 b^2 b^\omega [(eF)^{-2}(1 - 
\tau eF \coth \tau eF)]_{\omega\omega'} b^{\omega'}\biggr\}.
\eea

In order to get the contribution to 
$$
iem \, \tr \gamma^\mu \langle x|(H^2+m^2)^{-1}|x'\rangle \, ,
$$ 
it is sufficient to retain linear in $F_{\mu\nu}$ terms in Eq.(25), and, 
among them, only those are relevant, which are of the order $\tau^2$ in the 
expansion of the exponential,
$$
\exp \biggl[ i\tau \biggl({1\over2}\sigma^{\omega\omega'} 
eF_{\omega\omega'}+ 2m\gamma^5\widehat{b}\biggr) \biggr]= 
\biggl[1 + i\tau \biggl({1\over2}\sigma^{\omega\omega'} 
eF_{\omega\omega'}+ 2m\gamma^5\widehat{b}\biggr) -$$
$$- {{\tau^2}\over2} \biggl({1\over2}
\sigma^{\omega\omega'} eF_{\omega\omega'}+ 2m\gamma^5\widehat{b}\biggr)^2 + 
\ldots\biggr].
$$
Thus we get that this contribution is finite in the limit $x'\to x$:
\bea
iem \, \tr \gamma^\mu \langle x|(H^2+m^2)^{-1}|x\rangle &=&
-{ie^2 m^2 \over2(4\pi)^2}\tr \biggl(\gamma^\mu \sigma^{\omega\omega'} 
\gamma^5 \widehat{b} +  
\gamma^\mu \gamma^5 \widehat{b} \sigma^{\omega\omega'} 
\biggr)\,F_{\omega\omega'}\, \int\limits_{-\infty}^{0} d\tau 
\e^{i\tau m^2} =\nonumber\\
&&= - {e^2\over2\pi^2} \,\widetilde{F}^{\mu\nu}b_\nu ,
\eea
coinciding with Eq.(18), and, consequently, yielding Eq.(19). 

As to the CPT-even part of the effective action, instead of quantity
\be
ie \, \tr \gamma^\mu \langle x|\widehat{\pi} (H^2+m^2)^{-1}|x'\rangle =
- {ie\over2}\, \tr \langle x| {\delta \over{\delta \pi_\mu}} 
\int\limits_{-\infty}^{0} {d\tau \over {\tau}} \, \e^{i\tau (H^2+m^2)}
|x'\rangle \, ,
\ee
let us consider corresponding effective lagrangian
\be
L^{\rm quant}_1 = - {i\over2} \int\limits_{-\infty}^{0}
{d\tau \over {\tau}} \,\tr \langle x|\e^{i\tau (H^2+m^2)}|x\rangle \, ,
\ee
which, with the use of Eq.(25), takes form 
\bea
\hspace{-.5cm} L^{\rm quant}_1 &\!\!\!\!=&\!\!\!\! 
 - {1\over{2(4\pi)^2}} 
\int\limits_{-\infty}^{0}
{d\tau \over {\tau}^3}\biggl[\det\biggl({\sinh \tau eF \over 
\tau eF}\biggr)_{\rho\rho'}\biggr]^{-{1\over2}} 
\tr \exp\biggl[i\tau \biggl({1\over2}\sigma^{\omega\omega'} 
eF_{\omega\omega'} + \nonumber\\ 
&&\hspace{-.5cm} + 2m\gamma^5\widehat{b} - 4 m^2 b^2 b^\omega [(eF)^{-2}(1 - 
\tau eF \coth \tau eF)]_{\omega\omega'} b^{\omega'} + m^2\biggr)\biggr].
\eea
Subtracting the qudratically divergent ($\int\limits_{}^{0}
{d\tau \over {\tau}^3}$) contribution of the free Dirac operator, we 
get logarithmically divergent quantity
\be                             
L^{\rm quant}_1 - L^{\rm quant}_1 \big|_{F_{\mu\nu}=0}=
{e^2\over48\pi^2} F^{\mu\nu}F_{\mu\nu} \int\limits_{-\infty}^{0}
{d\tau \over {\tau}} \, \e^{i\tau m^2},
\ee
which yields effective action, 
$$
\Gamma^{\rm quant}_{\rm even}=\int
d^4x \, (L^{\rm quant}_1 - L^{\rm quant}_1 \big|_{F_{\mu\nu}=0})\,  ,
$$ 
coinciding with Eq.(23) under substitution $\int\limits_{-\infty}^{0}
{d\tau \over {\tau}} \, \e^{i\tau m^2} \to  \ln(tm^2)+\gamma$.

Thus, we conclude that the use of two methods -- the
one involving heat kernel coefficients and the one of proper time --
yieds consistent results; an obvious advantage of the heat kernel method is
that it allows one to get the results easily in the case of the
inhomogeneous field strength.

Taking, instead of $\Gamma$ (1), classical action without the 
Chern-Simons ($A\widetilde{F}$) term, as an input,
 \be
 \Gamma^{\rm class}= -{1\over4}\int d^4x\,F^{\mu \nu}(x)F_{\mu\nu}(x),
 \ee 
and adding $\Gamma^{\rm quant}_{\rm odd}$(19) and 
$\Gamma^{\rm quant}_{\rm even}$(23), we get total action, 
classical plus quantum,
\bea
\Gamma^{\rm class}+\Gamma^{\rm quant}&=& -
{1\over4}\biggl\{1 -{e^2\over12\pi^2}[\ln(tm^2)+\gamma]\biggr\}\int
d^4x\, F^{\mu\nu}(x)F_{\mu\nu}(x)-\nonumber\\
&&-{e^2\over4\pi^2} \,\, b^\mu \, \int
d^4x\,A^\nu(x)\widetilde{F}_{\mu\nu}(x).  
\eea 
The coefficients before the Maxwell $(F^2)$ term are gathered in a 
familiar manner to exhibit charge renormalization, 
\be
e^2_{\rm ren}=e^2\biggl\{1-{e^2\over12\pi^2}[\ln(tm^2)+\gamma]\biggr\}^{-1},
\ee
$e$ is the bare charge which is divergent as $t\to0_+$, and $e_{\rm ren}$
is the finite physical charge. Appropriately, electromagnetic field is 
renormalized,
\be
A^{\rm (ren)}_\mu(x) ={e\over e_{\rm ren}} \, A_\mu(x)\, , \quad
F^{\rm (ren)}_{\mu\nu}(x)={e\over e_{\rm ren}} \, F_{\mu\nu}(x),
\ee
and Eq.(32) takes final form
\be
\Gamma^{\rm class}+\Gamma^{\rm quant}= -
{1\over4}\int d^4x\,F^{\mu\nu}_{\rm (ren)}(x)F^{\rm (ren)}_{\mu\nu}(x)-
{e^2_{\rm ren}\over 4\pi^2} \,\, b^\mu \, \int
d^4x\,A^\nu_{\rm (ren)}(x)\widetilde{F}^{\mu\nu}_{\rm (ren)}(x) \, .
\ee
As a result, we have arrived at the same expression as Eq.(1), where, 
instead of $k^\mu$, stands 
\be
k^\mu_{\rm eff} = {e^2_{\rm ren}\over 4\pi^2} \, b^\mu \, .
\ee

It should be noted that operator $H^2$ (6) is nonhermitian in the case of 
timelike $b^\mu$. Perhaps, this reflects the tachyonic instability emerging 
in a theory governed by action (1) in the case of timelike $k^\mu$ 
\cite{Ca, AS}.

One might think that stringent astrophysical bounds 
\cite{Go, Ja1} should be imposed on the value of $k^\mu_{\rm eff}$ (36). 
However, the following circumstances have to be taken into account, 
warning against such a conclusion.

Firstly, there is no principle that rules out an appearance of the
($A\widetilde{F}$) term in the classical electromagnetic action. Thus, 
if we take $\Gamma$ (1) as a classical action, then, after adding quantum 
correction in the one-loop approximation, Eqs.(19) and (23), we arrive at 
\be
\widetilde{k}^\mu = {e^2_{\rm ren}\over e^2} \, k^\mu + 
{e^2_{\rm ren}\over 4\pi^2} \, b^\mu \, ,
\ee
instead of Eq.(36). Evidently, a restriction, if any, on the value of 
$\widetilde{k}^\mu$ leaves the value of $k^\mu_{\rm eff}$ (36) unrestricted, 
owing to indefiniteness of the value of $k^\mu$ and charge renormalization.
This is one of the sources of ambiguity in the value of the coefficient before 
the $A\widetilde{F}$ term, as discussed in the literature \cite{Pe}.

Secondly, the nature of constant vector $b^\mu$ has not been yet specified. 
A plausible assumption is that it is generated as the vacuum expectation 
value of a certain gauge vector field coupled to the axial-vector fermionic 
current. Then, if one sums over all fermion species (e.g., leptons and quarks 
in the standard model), the net result is zero \cite{Co2}, as a consequence of 
the axial anomaly cancellation condition which should be valid in the standard 
model and in any of its reasonable extensions as well. The individual values 
in each fermion sector are therefore irrelevant for comparison with 
astrophysical observation data. As it is shown in the present Letter, 
the use of the covariant nonperturbative formalism allows us to determine 
these individual values in the one-loop approximation unambiguously.
 
\medskip

\noindent{\bf Acknowledgements}

I am grateful to R. Jackiw for enlightening discussions and helpful remarks. 
I would like also to thank V.P. Gusynin, A.U. Klimyk, D.G.C. McKeon and 
M. Perez-Victoria for interesting discussions. The work was supported by 
U.S. Civilian Research 
and Development Foundation (CRDF grant UP1-2115) and Swiss National Science 
Foundation (SCOPES 2000-2003 grant 7IP 62607).


\end{document}